# Chapter 10
# Rybu: Imperative-style Preprocessor for Verification of Distributed Systems in the Dedan Environment

Wiktor B. Daszczuk [a,1], Maciej Bielecki[2] and Jan Michalski[2b]
[a] *Warsaw University of Technology, Institute of Computer Science*
[b] *Warsaw University of Technology, Faculty of Electronics and Information Technology*

**Abstract.** Integrated Model of Distributed Systems (IMDS) is developed for specification and verification of distributed systems, and verification against deadlocks. On the basis of IMDS, Dedan verification environment was prepared. Universal deadlock detection formulas allow for automatic verification, without any knowledge of a temporal logic, which simplifies the verification process. However, the input language, following the rules of IMDS, seems to be exotic for many users. For this reason Rybu preprocessor was created. Its purpose is to build large models in imperative-style language, on much higher abstraction level.

## 1. Introduction

In the education in computer science, one of important skills concerns parallel programming and synchronization, especially in distributed environment. Yet, there are difficulties in teaching this issues in undergraduate studies, where students do not have sound experience and mathematical basis. Students do not have proper sense of importance of proper synchronization, and teachers do not have the tools supporting proper evaluation of students' work. Without such tools, it is hard to track a parallel program and show errors or inconsistences in student's solution.

There are many methods for concurrent systems verification, including proving methods, bisimulation and model checking. Static methods are used in design phase, while dynamic ones during run-time. For evaluation of students' work, static methods are mostly applicable, but sometimes a possibility to simulate a program is very helpful. Yet, most verification formalism, like Petri net analysis or model checking technique requires some knowledge from the user, not typically possessed in early stages of studies in computer science. For the same reason, specification formalism like CSP [1] or CCS [2] may be introduced rather on older years of study.

---

[1] Corresponding Author. Nowowiejska Str. 15/19, 00-665 Warsaw, Poland, E-mail: wbd@ii.pw.edu.pl

Formal verification is widely used in computer systems development, however, the authors' experience with formal methods showed that designers are willing to use the results of formal verification carried out by another person, but they are reluctant to learn a formalism (for example, model checking). Many designers, especially students of first years of study, are not prepared to use advanced techniques, for example:

"… those who taught mathematical modeling (or "formal methods") faced daunting challenges. First, most modeling tools used seemingly esoteric notations that were hurdles for many students …" [3]

Formal methods give significant help if they may be used in automatic way, not requiring specialistic knowledge from the users. Automatic verification tool Alloy was built in MIT for the purpose of verification in student laboratory [4]. In many verifiers, automatic deadlock detection concerns total deadlocks, in which all processes participate [5]. In other approaches, partial deadlocks may be automatically checked, but only in systems specific process structure [6,7]. Alternatively, partial deadlocks may be identified using temporal formulas related to individual features of a model [8].

In Institute of Computer Science, WUT, a system Dedan (Deadlock analyzer, [9]) for modeling and automatic verification of distributed systems is used in student laboratory. Students model their solutions in IMDS (Integrated Model of Distributed Systems [10,11]) formalism and check them against deadlocks. Although Dedan is based on temporal logic and model checking, universal formulas are used to find deadlocks and the process of temporal verification is hidden inside the tool. The universality of formulas lays in the fact that they do not depend on the structure of the model. Both total and partial deadlocks may be identified. A user simply programs a solution in Dedan input format and checks the correctness in "push the button" manner. Dedan confirms the deadlock freeness or gives a counterexample in readable form of sequence diagram-like graph. A counterexample is a sequence of states of servers and messages sent between them, so the analysis of the counterexample helps in locating a source of error.

The Dedan tool offers additional analysis features like observation of the total graph of a verified system, simulation of the system over this graph, over a counterexample or over individual components of the system, where the components are converted to automata-like graphs.

The disadvantage of Dedan verification environment is the necessity of specification of a model in Dedan input language. This language is derived from the formal definition of IMDS, using sets of *servers*, *agents* (distributed computations), servers' state *values* and servers' *services*. Behavior of a system is defined as sets of actions "glued" by servers' *states*, or alternatively by or agents' *messages*. Actions grouped in one of these two manners form the processes: a server process contains all actions of this server (with states of this server on input) wile an agent process contains all actions of this agent (with messages of this agent in input). Thus, this specification style is far from typical programming style and is exotic to students.

Two authors of this paper (Maciej Bielecki and Jan Michalski), students of ICS WUT, developed a preprocessor called Rybu [12], which overcomes this drawback, and simplifies the specification using imperative programming-like input language ([13]-Ch. 2), with its syntax similar to a subset of C language. What is more, the language of Rybu has much higher level of abstraction and is able to generate large models from simple specifications due to Cartesian products of values used in servers source code. It is impossible to build such large models manually.

The elaboration of higher-level programming language for distributed systems specification, and Rybu preprocessor for Dedan environment, are contributions of this paper.

The paper is organized as follows: Section 2 contains an overview of IMDS formalism. Section 3 presents an example of a verification under Dedan. In Section 4 main ideas of Rybu are presented. Rules of conversion of Rybu constructs to Dedan input are given in Section 5. Section 6 contains an example of advanced verification using Rybu. Conclusions and future directions of Dedan and Rybu development are covered in Section 7.

## 2. Integrated Model of Distributed Systems

Typically, a distributed system is modeled as a set of *servers* exchanging messages. Such a view follows a *client-server* paradigm [14]. On the other hand, processes traveling between servers are used in *remote procedure calling* paradigm (RPC [14]). Such travelling processes will be called *agents* in the paper, and a called procedure will be called a *service*, which is a typical term in RPC view. In the Integrated Model of Distributed Systems both views are included in uniform formalism.

The servers' states are the pairs (*server, value*). Also, the servers offer sets of services which may be invoked by the agents' messages. Therefore the messages are triples (*agent, server, service*).

The system behavior is built over *actions*, which have a form of relation (*message, state*) $\lambda$ (*new_message, new_state*). A special kind of action terminates an agent: a new message is absent. The actions are executed in interleaving manner [15]. The processes are extracted from the specification of a system: *server processes* group the actions with common server's states on their input (and on output as well). *Agent processes* group the actions with common agent's messages on their input (and on output, if present). There are two possible decompositions of a system: a decomposition into server processes is the *server view*, while decomposition into agent processes is the *agent view* [10].

A *configuration* of the system consists of all servers' states and all agents' messages (except terminated ones). An *initial configuration* consists of *initial states* and *initial messages*. The semantics of a system is defined by a Labeled Transition System (LTS, [16]). Configurations are the nodes of LTS (initial configuration is the initial node) and actions are the transitions of LTS. The formal definition of IMDS is given in the Appendix, and it is described in [11].

The important feature of IMDS is asynchronous specification, which it exploits natural features of distributed systems:

- *locality* of action in servers: any action of a server is executed on a basis of current state of this server and messages pending at this server, no notion like global state agreed between servers may enable of prohibit any action (which is the case in synchronous formalisms like CSP [1] or CCS [2]),
- *autonomy* of servers: a server itself decides which one of the enabled actions will be executed first, no external element can influence such decisions,
- *asynchrony of actions*: a server's state waits for matching messages or a pending message waits for a matching state, no elements occur in synchronized way,

- *asynchrony of communication*: messages are passed over unidirectional channels between servers without any synchronization of sending a message along a channel with receiving it; the only way to send a reply is to send a response message over a second, independent channel in reverse direction; synchronous models like CSP or CCS assume some knowledge about global state of a system to perform inter-process communication, which is unrealistic in distributed systems.

In Lotos, asynchronous specification is used, but verification typically concerns total deadlocks [17]. Specification in IMDS allows for automatic partial or total deadlock detection in a modeled system. It is achieved by using universal (not related to a structure of verified system) temporal formulas over IMDS elements: states and messages [11].

**3. Example verification in Dedan – two semaphores**

The example of deadlock detection is presented on the system of two threads *proc1* and *proc2* using two semaphores *sem1* and *sem2*. The agents come from their own servers, and each semaphore has its own server. The pseudo-code of the system is:

```
proc1:          proc2:
sem1.wait;      sem2.wait;
sem2.wait;      sem1.wait;
sem1.signal;    sem2.signal;
sem2.signal;    sem1.signal;
stop            stop
```

This system falls into a total deadlock when *proc1* holds *sem1* and waits for *sem2* and *proc2* holds *sem2* and waits for *sem1*. The system is presented below in IMDS notation, which is the input notation of Dedan. The specification is in the server view. It is simply a grouping of actions on individual servers. A system is defined as a sequence of server type specifications (enclosed by **server** … *};* – lines 2-9, 10-19*)*, server and agent instances (variables) declaration (**agents** …, **servers** … – lines 20,21) and an initial configuration phrase (**init** → *{...}* – lines 22-27). A server type heading (l. 2,10) contains a set of formal parameters: agents and servers used in actions of the server type. Formal parameters may be vectors, as *A[2]* and *proc[2]* (l.2). Each server has a set of services (l.3,11), a set of states (l.4,12) and a set of actions assigned (in arbitrary order, l.6-8, 14-18). Services and states may be vectors. For a compact definition, repeaters may precede the actions in a server type definition (l.6-8). A repeater is an integer variable with defined range. The definition of an action is repeated for every value of a preceding repeater. If many repeaters are applied, a Cartesian product of their values is used. The indices of agents, states and services indicate individual instances (l.6-8).

Server and agent variables may be organized in vectors (l.20,21). In initialization part (l.22), actual parameters are bounded to formal parameters (again, repeaters and indices may be used – l.23-26). The initial states of servers and the initial messages of agents are also assigned.

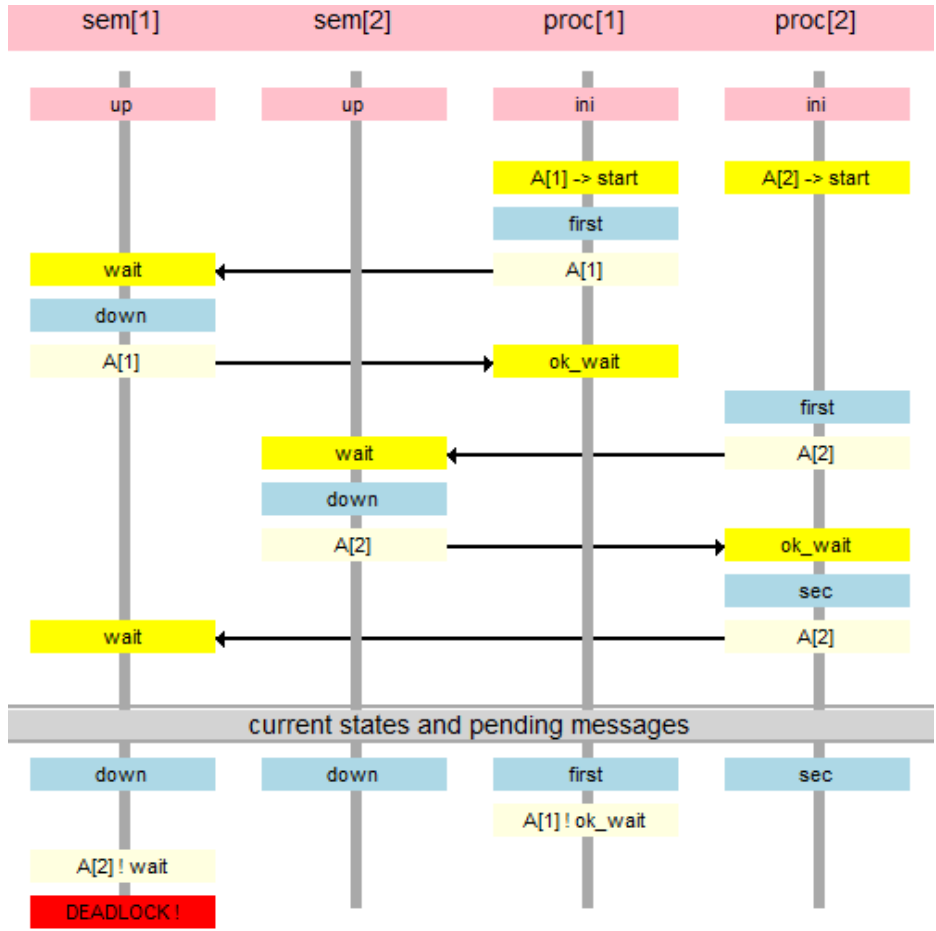

**Figure 1.** The counterexample for a deadlock found in the two_sem system.

```
1. system      two_sem;

2. server:     sem (agents A[2]; servers proc[2]),
3. services    {wait, signal},
4. states      {up, down},

5. actions     {
6. <j=1..2>    {A[j].sem.wait, sem.up} -> {A[j].proc[j].ok_wait, sem.down},
7. <j=1..2>    {A[j].sem.signal, sem.down} -> {A[j].proc[j].ok_sig, sem.up},
8. <j=1..2>    {A[j].sem.signal, sem.up} -> {A[j].proc[j].ok_sig, sem.up},
9. };
```

```
10. server:      proc (agents A; servers sem[2]),
11. services     {start, ok_wait, ok_sig},
12. states       {ini, first, sec, stop},
13. actions      {
14.              {A.proc.start, proc.ini} -> {A.sem[1].wait, proc.first},
15.              {A.proc.ok_wait, proc.first} -> {A.sem[2].wait, proc.sec},
16.              {A.proc.ok_wait, proc.sec} -> {A.sem[1].signal, proc.first},
17.              {A.proc.ok_sig, proc.first} -> {A.sem[2].signal, proc.sec},
18.              {A.proc.ok_sig, proc.sec} -> {proc.stop},
19. };

20. agents:      A[2];
21. servers:     sem[2], proc[2];

22. init ->      {
23. <j=1..2>     A[j].proc[j].start,
24.              proc[1](A[1],sem[1],sem[2]).ini;
25.              proc[2](A[2],sem[2],sem[1]).ini;
26. <j=1..2>     sem[j](A[1],A[2],proc[1],proc[2]).up,
27. }.
```

The verification is performed in the Dedan environment. An obvious deadlock is found, which counterexample is shown in Fig. 1.

## 4. Overview of the Rybu preprocessor

The main purpose of Rybu is to simplify verification in Dedan. Therefore, a limited class of systems may be modeled, but the usage should be more convenient compared to IMDS notation. Namely, modeled systems are organized within client-server paradigm [14]. Other kinds of systems, for example modeling vehicle guidance systems, where servers model road segment controllers and agents model the vehicles [18], cannot be specified in Rybu.

A basis of Rybu is the server view of a modeled system. A Rybu system consists of two kinds of servers: reactive servers, similar to the semaphores in the example, which serve as resources and offer their states and services to other servers. From now on, the term *server* will be used in the meaning reactive servers. If we mention a server in IMDS formalism, it is named *IMDS server*.

The second kind of servers performs distributed computations, called *threads*. The agents origin and run on threads, calling the services offered by the servers.

### 4.1. Servers

The services of servers are invoked by threads by means of messages. A server consists of *state variables* and *actions*. State variables are implemented as sets of IMDS servers' *values*, and actions are implemented as sets of IMDS actions. Actions define how the invocations of the servers' services, which may be regarded as procedure calls from the threads, are executed. When such a service is called, the server changes the values of its state variables and returns a response (being a message itself). *Return values* – technically threads' resources – are atoms of an enumeration type bound to specific action, which is defined by usage.

Rybu action is a 4-tuple of (*service*, *predicate*, *return value*, *state update*) where:

- *service* – an identifier used by a thread to trigger a server's action,
- *predicate* (optional) – a condition (over state variables) on which the action may be executed,
- *return value* – a value returned to the calling thread,
- *state update* (optional) – a function mapping a server's state to a new state of the server, represented in the syntax as assignment to one or more state variables.

The server type `sem` from the `two_sem` example has the following form in Rybu:

```
1. server sem {
2.   var state : {up, down};
3.   { wait | state == :up } -> { state = :down; return :ok; }
4.   { signal } -> { state = :up; return :ok; }
5. }
```

The service `wait` may be executed on condition that the state of the semaphore is `:up`, then it is updated to `:down` and the response `:ok` is issued to the caller. The service `signal` may be executed any time, it leaves the state of the semaphore `:up` and confirms it by the `:ok` response.

*4.2. Actions*

Each action (*service*, *predicate*, *return value*, *state update*) is applied to a *service*, for the value of *state variables* for which the *predicate* evaluates to true. If the predicate is true, the action causes state variables to receive the values appointed by *state update* function and a response if issued with *return value* to a calling *thread*.

Note that for a specific *service*, more than one *action* may be defined, with different *state updates*, and intersecting set of *state variables* values. This causes nondeterministic state transitions of *state variables*. In this case the actions are nondeterministic.

*4.3. Threads*

The actions in IMDS server are not set up in any sequence, it is only a set of actions and it may be listed in arbitrary order. It is a case because there is no notion comparable to a "program counter", which causes the instructions to be executed in a sequence. In Rybu, the instructions are threaded in the order of placement as in typical imperative programming language ([13]-Ch.2). Additional structural instructions may change the order of execution, for example *loop* statement or choice (*match*) statement. The IMDS server `proc[1]` with its agent `A[1]` are modeled in Rybu as:

```
1. var sem1 = sem() { state = :up };
2. var sem2 = sem() { state = :up };
3. thread proc1() {
4.    sem1.p();
5.    sem2.p();
6.    sem1.v();
7.    sem2.v();
8. }
```

Variables *sem1* and *sem2* are instances of the sever type *sem*. The thread *proc1* calls the services *p* and *v* of the server type *sem* in the order just taken from the pseudo-code at the beginning of Section 3. The IMDS server *proc[2]* with its agent *A[2]* is very similar, except that the order of operations on the semaphores is switched (*sem2*, then *sem1*).

*4.4. Statements*

The Rybu preprocessor uses single executive statement **match**. It consists in sending a message to a server (calling its service) and waiting for response (reply message). Then, a response value may direct the control to a proper follow-up statement. The *server.service();* statement form, used in Section 4.3, is a simplified version. It is used when the service returns only *:ok* response value. The example of the full version of **match** statement is contained in Section 5.3.

There is also the control statement **loop** *{… statements …}*.

*4.5. Communication in Rybu*

Communication in Rybu follows the scheme of a remote procedure call [14]: a service of a server is invoked by a thread, then it waits for a response inactively. The response may contain a value of an enumeration type.

*4.6. Data types*

The set of possible server states is the Cartesian product of ranges of all of its *state variables*. Data types are necessary to determine the range of possible state variable values. Each type denotes a finite set of values. There are three types available:

1. Integer range: *min..max* (where *min* and *max* are integer-valued expressions) -- values of this type are just integers in range [*min; max*].
2. Enumeration: *{atom1, atom2}* - values of an enumeration are single atoms, e.g., *:atom1*. Note that the atoms are preceded by a colon when used.
3. Vectors of the above types: *elem_type[length]* - values are concrete vectors of values of *elem_type*, holding the *length* size, e.g., for the vector type *(0..3)[4]*, a possible vector value could be *[0,3,1,2]*.

*4.7. Organization of Rybu*

The program consists of three main parts:

1. *Parsing*. A recursive descent parser is used, implemented using Monadic Parser Combinators [19]. This process generates an Abstract Syntax Tree [20].
2. *Conversion*. Rybu servers and threads are converted to state machines, represented by sets of state transitions.
3. *Output generation*. The resulting automata are printed as IMDS servers in Dedan syntax.

The implementation in Haskell is covered by a suite of unit tests and a set of (input file, expected output) pairs. The latter can be expensive to maintain, as the test result is sensitive to the ordering of actions produced by the program (which has no semantic meaning).

## 5. Conversion rules

In this section, implementation all constructs of Rybu in IMDS is presented: servers, actions and variables in servers, threads and messages.

*5.1. Servers and servers' variables*

Each Rybu *server* is converted to an IMDS *server type*. The set of its IMDS *states* is defined by a Cartesian product of sets of all possible values of the *state variables* declared in the server.

For example, Rybu server `test`:

```
1. server test {
2.   var val1: 1..3;
3.   var val2: {true, false};
4. }
```

generates an IMDS server that has the following set of *states*:
{`val1_1_val2_true, val1_1_val2_false, val1_2_val2_true, val1_2_val2_false, val1_3_val2_true, val1_3_val2_false`}
Proper server instances are declared, and in the initialization part, state variables are initialized, for example for the declaration:
  `var t = test() {val1 = 1; val2 = :false};`

the IMDS declaration is:
  `servers t:test,…`

and the IMDS initialization is:
  `t(`*… used servers and agents …*`).val1_1_val2_false,`

*5.2. Actions in server*

An action (*service*, *predicate*, *return value*, *state update*) is converted to a set IMDS actions applied to a *service*, for all possible values of *state variables* for which the *predicate* is true. For every pair of *(service, state variables values)*, for which the predicate is evaluated to true, an IMDS action is generated with the following properties:

- The IMDS *input state* is the one of which the predicate was evaluated to true.
- The IMDS *input message* calls the *service*.
- An *output state* is obtained by application of the *state update* function to the input state (state variables mentioned get their new values while all other state variables preserve their values).
- The above two steps are applied to each IMDS *agent* implementing a *thread* that invokes the service (see Section 5.3 Treads).
- A *return value* is sent and IMDS *output message* to the IMDS server implanting the *thread* which calls the action.

For example, the action 4 in the example in Section 4.1 is converted to a following set of IMDS actions, provided that two threads `t1` and `t2` invoke the service `signal`. The example thread `t1` is implemented as IMDS server `S_t1` and its agent `A_t1`, similarly `t2`.

```
{A_t1.sem.signal, sem.up} -> {A_t1.S_t1.ok, sem.up},
{A_t1.sem.signal, sem.down} -> {A_t1.S_t1.ok, sem.up},
{A_t2.sem.signal, sem.up} -> {A_t2.S_t2.ok, sem.up},
{A_t2.sem.signal, sem.down} -> {A_t2.S_t2.ok, sem.up},
```

*5.3. Threads*

Each *thread* is converted into an IMDS server and an accompanying agent. A *thread* issues messages invoking some servers' services and then it waits for a response corresponding to this call. States of a thread are defined by the "position in code" (a virtual "program counter"). From each such state there are as many IMDS actions generated as the number of branches in the `match` statement, each reacting to a specified response. Below an example of Rybu code shows thread syntax with **match** statement which branches depending on response received from calling the service `y` on the server `s1`.

```
1. thread x() {
2.     loop {
3.             match s1.y(){
4.                 :ok => s2.z();
5.                 :er => s3.v();
6.         }
7.     }
8. }
```

In IMDS, every statement is associated with its own value of an IMDS server's state, which acts as a „program counter". The above example is converted to a set of IMDS actions:

```
    {A_x.S_x.ok, S_x.s0_s1_y} -> {A_x.s2.z, S_x.s2_s2_z},
    {A_x.S_x.er, S_x.s0_s1_y} -> {A_x.s3.v, S_x.s5_s3_v},
    {A_x.S_x.ok, S_x.s2_s2_z} -> {A_x.s1.y, S_x.s0_s1_y},
    {A_x.S_x.ok, S_x.s5_s3_v} -> {A_x.s1.y, S_x.s0_s1_y},
```

States *s0_s1_y, s2_s2_z, ...* are used as a „program counter". Every IMDS server implementing a thread starts from *s0_...* state. In IMDS initialization part, the IMDS server is initialized and its agent executes its first message in the thread's code:

```
    S_x(s1,s2,s3,A_x).s0_s1_y,
    A_x.s1.y,
```

*5.4. Communication*

The communication in Rybu follows the remote procedure call principle [14]. A call consists of two messages: one to call a server and the other one that carries a response. A thread calls a server by issuing a message that invokes the server's service (for example *A_x.s2.z* above). A response is a message in the opposite direction, sent from the server to the calling thread, like *A_x.S_x.er*. The pair of messages is passed in the context of the thread's agent *A_x*.

## 6. Advanced verification in Rybu

Consider a system consisting of two bounded buffers. The buffers are modeled by the counters representing current numbers of elements – the contents are irrelevant for synchronization purposes.

There are two threads continuously trying to balance the buffers – they move items from a buffer that is more filled to the other one. The threads are symmetric, but every thread checks different direction of moving elements.

A deadlock happens because a check and an action are not guarded by a critical section, for example when *shouldPut1* check causes the *put2* action. This allows for interleaving that may cause *shouldPut* checks to happen in both user threads first, and then both threads *put* elements to the same buffer, but there is a room only for one element. In this case both threads block on the same semaphore waiting for an element to appear in the other buffer, but there is no third thread running to put an element there: a deadlock occurs. The following code shows an implementation of this model in Rybu.

```
const N=3;

server SemN {
    var value: 0..N;

    { signal | value < N } -> { value = value + 1; return :ok; }
    { signal | value == N } -> { return :ok; }

    { wait | value > 0 } -> { value = value - 1; return :ok; } }
}
```

<’

```
server Buf {
    var count1: 0..N;
    var count2: 0..N;

    { shouldPut1 | count1 - count2 <= 0 } -> { return :true; }
    { shouldPut1 | count1 - count2 > 0 }  -> { return :false; }

    { shouldPut2 | count2 - count1 <= 0 } -> { return :true; }
    { shouldPut2 | count2 - count1 > 0 }  -> { return :false; }

    { put1 | count1 < N } -> { count1 = count1 + 1; return :ok;}
    { get1 | count1 > 0 } -> { count1 = count1 - 1; return :ok;}
    { put2 | count2 < N } -> { count2 = count2 + 1; return :ok;}
    { get2 | count2 > 0 } -> { count2 = count2 - 1; return :ok;}
}

var buf = Buf() { count1 = 0, count2 = 0 };
var sBuf1 = SemN() { value = 0 };
var sBuf2 = SemN() { value = 0 };

thread User1() {
    loop {
        match buf.shouldPut1() {
            :true  => {
               buf.put1();
               sBuf1.signal();
               sBuf2.wait();
               buf.get2();
            }
            :false => {
               buf.put2();
               sBuf2.signal();
               sBuf1.wait();
               buf.get1();
            }
        }
    }
}

thread User2() {
    loop {
        match buf.shouldPut2() {
            :true  => {
               buf.put2();
               sBuf2.signal();
               sBuf1.wait();
               buf.get1();
            }
```

```
            :false => {
               buf.put1();
               sBuf1.signal();
               sBuf2.wait();
               buf.get2();
            }
         }
      }
}
```

The example illustrates the use of multiple state variables (two range variables in the *Buf* server) and imperative programming of the threads.

The deadlock may be fixed by introducing a mutual exclusion semaphore that allows only one thread at a time to perform *shouldPut* check and *put* operation in a critical section.

## 7. Conclusions and future work

The verification of synchronization projects, performed by the students, prove the usefulness of verification in the Rybu/Dedan environment. Students appreciate automaticity of verification, which does not require from them any knowledge on temporal logics and model checking.

Rybu allows to prepare larger models, and on higher level of abstraction. Many cases were checked, including large ones, impossible to model directly in Dedan. For example, 131 lines of Rybu code of a student's solution generates about 6,000 of Dedan code. Till now, no student project exceeding available memory occurred. However for such a possible case, export of IMDS specification to external verifiers using reachability space reduction was prepared (for example Uppaal [21]).

Some other examples were verified, they may be found in [22]. Also, a verification of a large model of the Karlsruhe Production Cell benchmark [23] was performed [24]. Unlike in other approaches, simple asynchronous protocols are used between distributed controllers of individual road segments [16] or Production Cell devices [23]. Such a modeling is suitable for Internet of Things approach (IoT [25,26]), where independent devices autonomously negotiate their common behavior using simple protocols. Rybu/Dedan couple allows for rapid prototyping approach of distributed solutions, because of automatic verification of prepared models.

Also, real-time constraints may be applied to action and communication time delay for verification of real-time systems [24]. Dedan is equipped with real-time constraints which may be imposed on actions and communication. Such constraints may be imported to Rybu.

Rybu is based on client-server paradigm, but other approaches are possible. For example, agents in the system may travel through the servers, not returning to the servers they origin from. This is a subject for further research concerning a different kind of input language, addressed to traveling agents.

In Rybu some improvements are needed, for example state variables in threads, thread types, enriched communication statements.

*W. B. Daszczuk et al. / Rybu: Imperative Preprocessor for Verification of Distributed Systems in Dedan*

## Appendix – IMDS formal definition

$S =$ $\{s_1, s_2, ...\}$ – a finite set of servers
$A =$ $\{a_1, a_2, ...\}$ – a finite set of agents
$V =$ $\{v_1, v_2, ...\}$ – a finite set of values
$R =$ $\{r_1, r_2, ...\}$ – a finite set of services
$P \subset$ $S \times V$ – a set of states
$M \subset$ $A \times S \times R$ – a set of messages
$H =$ $P \cup M$ – a set of items
$T \subset$ $H$;
  $\forall_{p,p' \in T \cap P}\ p = (s,v), p' = (s',v'),\ p \neq p' \Rightarrow s \neq s'$;
  $\forall_{m,m' \in T \cap M}\ m = (a,s,r),\ m' = (a',s',r'),\ m \neq m' \Rightarrow a \neq a'$ – a configuration; one state for every server, at most one message for every agent
$T_0 \subset$ $H$;
  $\forall_{s \in S} \exists_{p \in T_0 \cap P}\ p = (s,v), v \in V$;
  $\forall_{a \in A} \exists_{m \in T_0 \cap M}\ m = (a,s,r),\ s \in S,\ r \in R$ – initial configuration; one state for every server, one message for every agent
$\Lambda \subset$ $(M \times P) \times (M \times P) \cup (M \times P) \times (P)\ |\ (m,p)\Lambda(m',p') \vee (m,p)\Lambda(p')$,
  $m = (a,s,r) \in M,\ p=(s_1,v_1) \in P,\ m'=(a_2,s_2,r_2) \in M,\ p'=(s_3,v_3) \in P,\ s_1 = s,\ s_3 = s,\ a_2 = a$
  – a set of actions – ordinary ones and agent-terminating ones[2]
$T_{inp},\ T_{out}$: $\forall\ _{\lambda \in \Lambda}\ \lambda = ((m,p),(m',p'))\ T_{inp}(\lambda) \supset \{m,p\},\ T_{out}(\lambda) = T_{inp}(\lambda) \setminus \{m,p\} \cup \{m',p'\}$ – obtaining $T_{out}(\lambda)$ from $T_{inp}(\lambda)$ for an action
$LTS =$ $< N, N_0, W >\ /$
  $N = \{T_0, T_1, ...\}$ (nodes);
  $N_0 = T_0$ (initial node);
  $W$ is the set of directed labeled transitions, $W \subset N \times \Lambda \times N$,
  $W = \{\ (T_{inp}(\lambda), \lambda, T_{out}(\lambda))\ |\ \lambda \in \Lambda\})\ \}$ – the labeled transition system

---

[2] Thus, $\Lambda$ is not strictly a relation, because it contains both quadruples and triples. More formally, a message „agent termination" may be added to $M$, which is prohibited on input of any action.

$B(s) =$   $\{\lambda \in \Lambda \mid \lambda = (((a,s,r),(s,v)), ((a,s',r'),(s,v'))) \vee \lambda = (((a,s,r),(s,v)), ((s,v'))),\ s' \in S,\ a \in A,$
   $v,v' \in V,\ r,r' \in R\ \}$ – server process of a server $s \in S$

$C(a) =$   $\{\lambda \in \Lambda \mid \lambda = (((a,s,r),(s,v)), ((a,s',r'),(s,v'))) \vee \lambda = (((a,s,r),(s,v)), ((s,v'))),\ s,s' \in S,$
   $v,v' \in V,\ r,r' \in R\ \}$ – agent process of an agent $a \in A$

**B** =   $\{B(s) \mid s \in S\}$ – the server view; a decomposition of a system into server processes

**C** =   $\{C(a) \mid a \in A\ \}$ – the agent view; a decomposition of a system into agent processes